\documentstyle[12pt]{article}

\include{psfig}
\def\beq{\begin{equation}}
\def\eeq{\end{equation}}

\begin{document}
\topmargin=-0.5in
\begin{center}
\thispagestyle{empty}
\bigskip

\vspace*{1in}
{\large \bf Self Organized Critical Dynamics of a Directed Bond Percolation Model} 
\bigskip

Subhankar Ray  {\scriptsize $^{^{\dag }}$}
\footnote{Present Address: Dept of Physics, Jadavpur University, Calcutta 700 032, India.\\ subho@juphys.ernet.in}, Tapati Dutta {\scriptsize $^{^{*}}$}, Jaya Shamanna {\scriptsize $^{^{\dag}}$} \footnote{Present Address: Institute of Physics, Bhubaneswar 751005, India.\\ shamanna@beta.iopb.stpbh.soft.net}

\noindent
{\scriptsize $^{^{\dag}}$ } Department of Physics, State University of New York \\ Stony Brook, New York 11794. \\
\noindent
{\scriptsize $^{^{*}}$} Physics Department, St. Xavier's College, Calcutta 700 016, India. \\

\bigskip

\end{center}
\bigskip
\begin{abstract}

\noindent
We study roughening interfaces with a constant slope that become
self organized critical by a rule that is similar to that of invasion percolation.
The transient and critical dynamical exponents show Galilean invariance.
The activity along the interface exhibits non-trivial power law
correlations in both space and time. 
The probability distribution of the activity pattern
follows an algebraic relation. \\

\noindent

\end{abstract}



\section{Introduction }

Self organized critical systems a la invasion percolation and interface
dynamics have generated good deal of interest in recent times 
[1 - 7].
Most of these systems show self affine or self similar structures. 
They show interesting time time power law correlation. 
The critical exponents of several of these systems were studied
in the said references.

We study a model of roughening interface proposed by Barma \cite{mbarma}
to investigate the dynamical exponents at the transient and saturation
regions of interface widths. The exponents for time time correlation of interface
width, and the time height correlations are found.
The time evolution of the activity center, the point at which the movement
begins, is determined. Unlike most other cases, there is no
`power law' and the distribution of `active center movement' with time
follows an algebraic relation, a linear relation with a negative slope.

The model is defined on a tilted square lattice where each bond (k)
on the lattice is assigned a quenched, uncorrelated random number $f_k$
drawn from the interval $[0,1[$. In the one dimensional version, a discrete
interface $h(x)$ is defined on a chain $x=1,2 \dots ,L$, $L$ being the
system size. We use cylindrical boundary condition. The chain is updated
by finding the smallest $f_k$ on the chosen interface. 
In order to preserve the length of the interface, the directed walk
character is maintained by local re-adjustments. If the chosen minimal
bond has a positive (negative) slope, the sequence of links with 
negative (positive) slope just below and on the left, also advance
as shown in figure 1. 

\def\thefigure{1}
\begin{figure}
\centerline{\psfig{figure=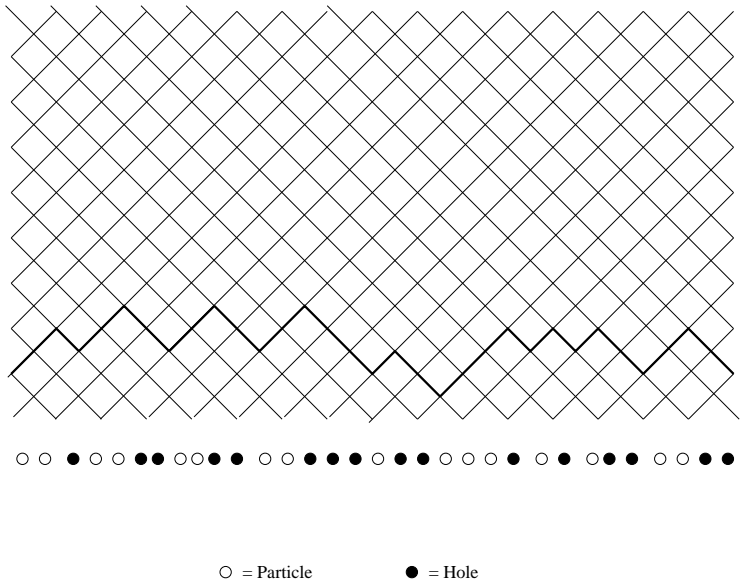,height=3in,angle=-90}}
\vspace{0.in}
\caption{Interface profile for the present model on a tilted lattice}
\label{fzm.ps}
\end{figure}

\def\thefigure{2}
\begin{figure}
\vspace*{-1in}
\centerline{\psfig{figure=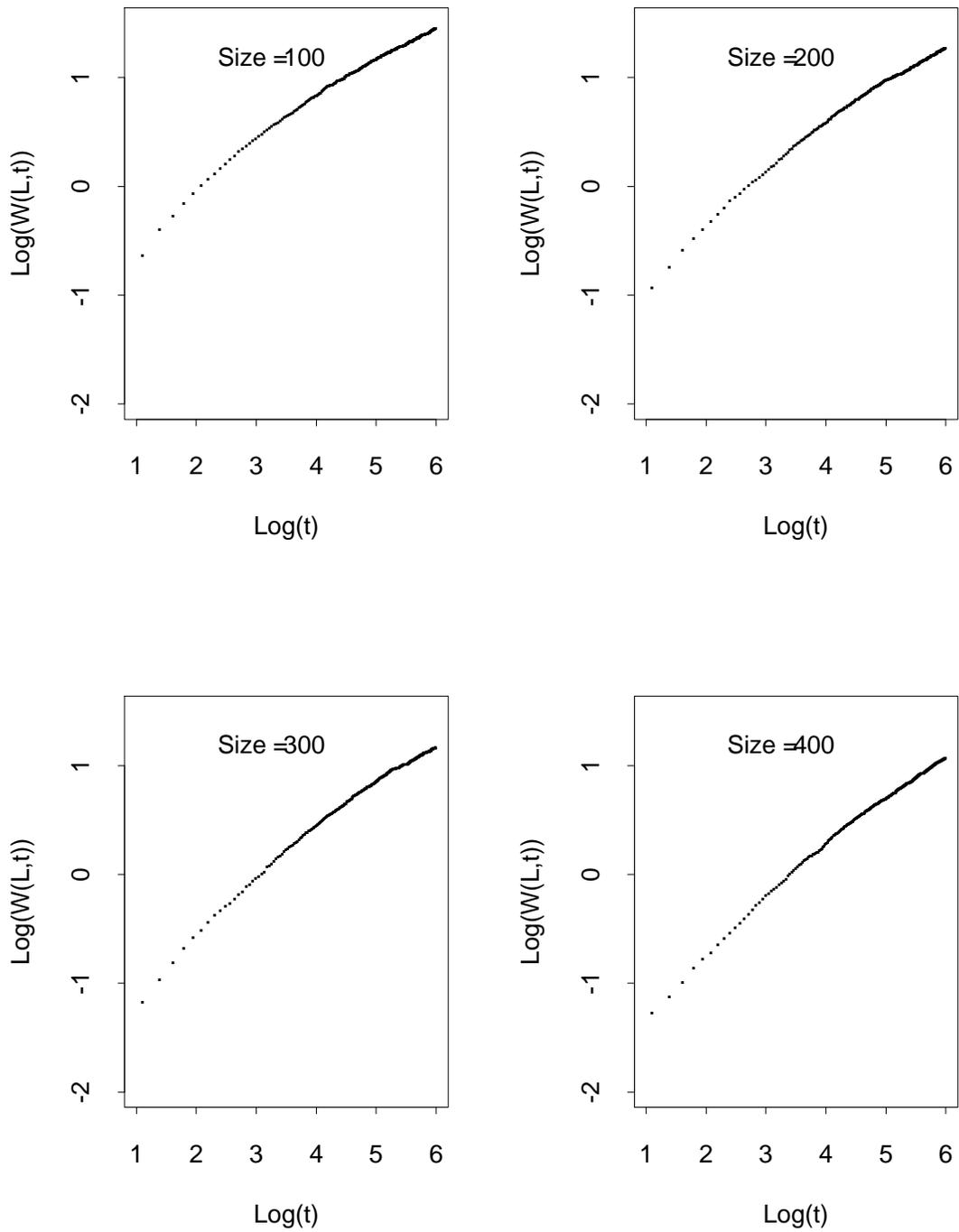,height=8in,angle=0}}
\vspace{0.in}
\caption{Variation of width W(L,t) with size L for different time step}
\label{fzm.ps}
\end{figure}

The interface dynamics is mapped onto a system of hard core particles
on a ring. A positive slope link of the interface is represented
by a particle ($n_k = 1$), and a negative slope link by a hole ($n_k = 0$),
see figure 1.
The difference in height of the interface between sites $J_1$ and $J_2$
is ;
  
\beq
h_{j_2} - h_{j_1} = \sum_{j=j_1}^{j_2} (2 n_j - 1)
\eeq

Each site $k$ on the ring carries a random number $f_k$ assigned
to the bond in the actual lattice. In each time step activity is
initiated at a site with a minimum $f_k$. If this site contains
a particle (hole) it exchanges place with the first hole (particle) on 
the left (right). A new set of $f_k$s are assigned to the particle 
and hole that are exchanged. All sites hopped over in this process
are also assigned new $f_k$s. The overall particle density $\rho$ is
conserved throughout the process. $\rho$ determines the mean slope
of the interface $m = 2 \rho -1 $.
Here progress of the interface is in the forward direction only.

\section{Results and Discussion}
\def\thefigure{3}
\begin{figure}
\centerline{\psfig{figure=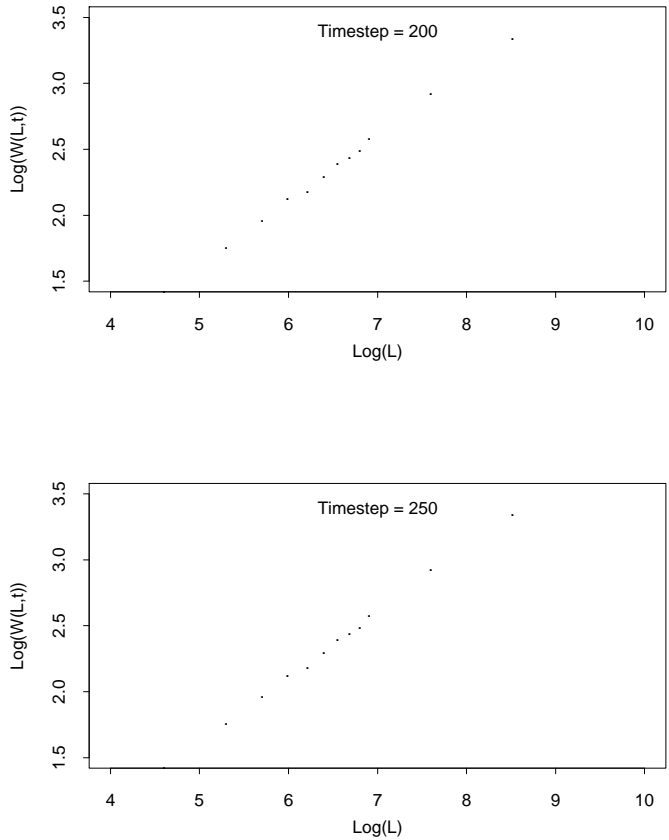,height=5in,angle=0}}
\vspace{0.in}
\caption{Variation of width W(L,t) with size L for different time step}
\label{fzm.ps}
\end{figure}

Sneppen \cite{sne92} investigated numerically the power law dependence 
of interface width with time t and system size $L$. 
We studied the interface width or the time time correlation of height defined
in terms of the standard deviation
\beq
W(L,t) = < ( [h(x,t+\tau) - h(x,\tau)] - <h(x,t+\tau) - h(x,\tau)> )^2 > ^{\frac{1}{2}}
\eeq
where $\tau$ determines the various starting times. The average $< \;\;\; >$
is done over $x \in [1,L]$ and members of the simulation ensemble.
We plot in figure 2 the temporal growth behaviour of the interface width,
the transient growth and the saturation region, for a surface starting from
a flat interface ($m=0$).
The saturation is identified in terms of saturation of slope.
The model shows different scaling behaviour in these two regions.

\def\thefigure{4}
\begin{figure}
\centerline{\psfig{figure=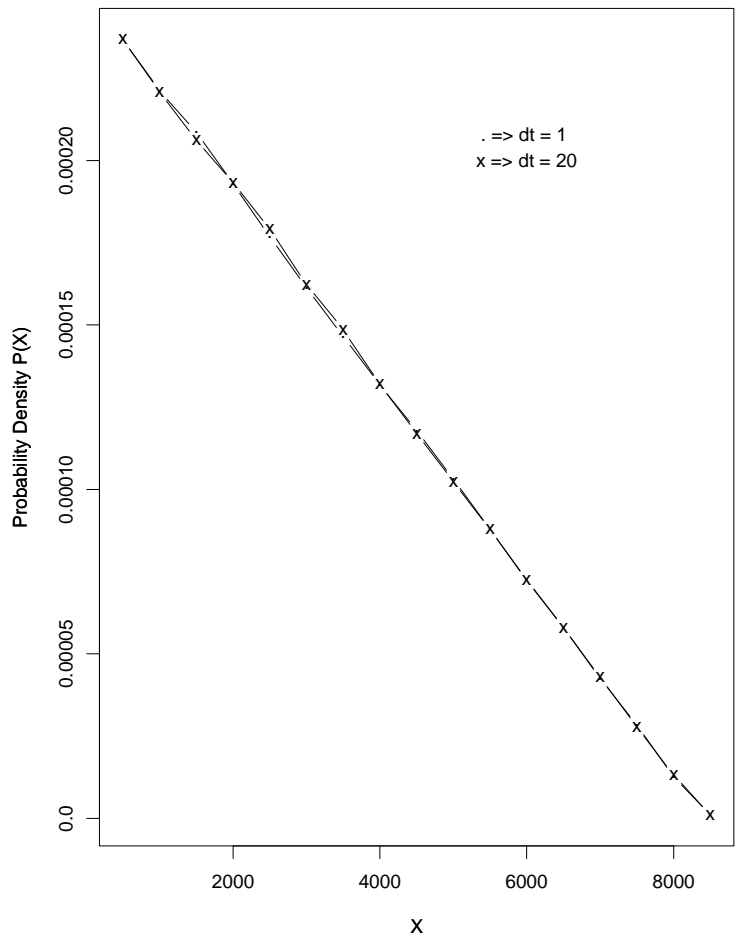,height=5.5in,angle=0}}
\vspace{0.in}
\caption{Probability distribution of active center displacement}
\label{fzm.ps}
\end{figure}

In table 1 we present the values of the transient and saturation exponents
of different lengths $L$. In the transient region we found the power law
behaviour 
\beq
W(L,t) \sim t^{\beta_{trans}} \;\;\;\;\;\; with \;\;\; \beta_{trans} = .4341 
\eeq
In the saturation region we get
\beq
W(L,t) \sim t^{\beta_{sat}} \;\;\;\;\;\; with \;\;\; \beta_{sat} = .3485
\eeq

The temporal behaviour of time time correlation of height in terms of 
the range is also studied in the literature
\beq
H(L,t) = < |max_x(h(x,t+\tau)-h(x,\tau)) - min_x(h(x,t+\tau)-h(x,\tau))| >_{\tau}
\eeq
The average $< \;\;\; >$ is done over $x \in [1,L]$ and members of the simulation 
ensemble.
However this model and its algorithm dictates its value to be almost
always a constant, namely 2.  

A proof goes like this: $h(x,t+\tau)-h(x,\tau)$ has integer values 
0 and 2, or in multiples of 2 if sufficient time steps are allowed.
However the maximum over a long string is almost always 2 (or its multiple if
time is large), the minimum is 0. Hence $H(L,t)$ is almost always 2
(or a multiple thereof if we are looking at large time differences).

At any time t in the development of the interface the width $w = < (h-<h>)^2>^{1/2}$
of a saturated interface scales with the system size
\beq 
w(L,t) \sim L^{\alpha}
\eeq
where $\alpha$ is the roughness exponent. This scaling is interpretable
as a self organization of interface towards a critical attractor. In this
model the attractor is a string of sites having high value of $f_k$s.
The important feature is that the temporal drift of the interface towards the
saturation state required no fine tuning of any parameter. Thus this model
may be viewed as an example of self organized criticality in one dimension.

The variation of $w$ with length $L$ in log-log scale
for several time steps is found, a representative 
part of which is reported in figure 3.
Table 2 shows the relevant results for $\alpha$ for 
two time steps. Mean $\alpha$ was found to be $.5$ to the first place of decimal.
This is in excellent agreement with the standard Kardar-Parisi-Zhang scaling $\alpha =1/2$.
We also find that the interface motion maintains the Galilean
invariance $\alpha + \alpha/\beta_{sat} =1.92 \sim 2$.

Since our model is a one dimensional interface with no overhangs,
the study of the distribution of activity along this interface is simple.
Activity is described in terms of events on a string. An
event begins by finding the minimum probability $f_k$
on an interface. Let $x(\tau)$ and $x(t+\tau)$ denote the positions
of the events on the interface at time $\tau$ and $(t+ \tau)$ respectively.
We calculate the probability distribution function $P(X_t)$ in the
variable $X_t = | x(\tau) - x(t+\tau) | $. In figure 4 this distribution
function is shown for different values of $X_t$.
The probability density depends algebraically on $X_t$.
This is in distinction with power law dependence found in other cases,
e.g., Sneppen and Jensen \cite{sne93}. \\

The computation was done on `Hayawatha', a Silicon Graphics workstation.
Authors would like to thank Prof. A.N. Basu and Prof. A.K. Sen for
meaningful discussions.

\vskip 1cm

\noindent
{\large Tables}

\vskip .5cm

\noindent
Table 1. Transient and saturation exponents of time time correlation of
width at different length scales. 

\hskip 7cm 1000 simulation steps \\

\begin{tabular}{|r|l|l|c|}  \hline
Length & $ \;\;\;\;\;\;\;\; \beta_{trans}$ & $\;\;\;\;\;\;\;\; \beta_{sat}$ & $<\beta>$ \\
 & $\tau =30 ; \;\; t= 120$ & $\tau =300; \;\; t=100$ & \\ \hline
100 & 0.348509412823739 & 0.289099642351271 & \\
200 & 0.395537573475355 & 0.320416123901019 & \\
300 & 0.417463014858253 & 0.28759592598634  & \\
400 & 0.432676615565537 & 0.361634369182171 & $ <\beta_{trans}>  = 0.434103$ \\
500 & 0.465140826654886 & 0.369840157779528 & \\
600 & 0.447692227261676 & 0.342850618167307 & $ <\beta_{sat}>  =  0.348462$ \\
700 & 0.44819146054204  & 0.352392062283316 & \\
800 & 0.432095028834515 & 0.335543843346805 & \\
900 & 0.446593525738461 & 0.437833260989437 & \\
1000 & 0.507132066113331 & 0.387414162344172 & \\ \hline
\end{tabular}

\newpage

\noindent
Table 2. Roughness exponent of interface width at two saturation times. 

\hskip 7cm 3000 simulation steps 

\hskip 7cm 100 - 1000 string length 

\begin{tabular}{|r|c|c|}  \hline
Time & $ \;\;\; \alpha$ & $ < \alpha > $ \\ \hline
& & \\
200 &  0.495378 & 0.494865 \\
250 & 0.494351 & \\ 
& & \\ \hline
\end{tabular}

\vskip 1cm

\noindent
{\large Figures}

\vskip 1cm

\noindent
Fig.1. Interface profile for the present model on a tilted square lattice \\
Fig.2. Temporal growth behaviour of width W(L,t) for different sizes \\
Fig.3. Variation of width W(L,t) with size L for different time step \\
Fig.4. Probability distribution of active center displacement \\

\end{document}